\documentclass[aps, twocolumn, prb, showpacs,preprintnumbers,amsmath,amssymb,superscriptaddress]{revtex4-1}
\usepackage{amsmath,amsfonts,amssymb,graphics,graphicx,epsfig,color,times,indentfirst,layout,hyperref,bm}
\usepackage{ulem}
\hypersetup{linktocpage,,colorlinks=true}

\newcommand{\bk}{{\bf k}}

\newcommand{\bx}{{\bf x}}

\newcommand{\bR}{{\bf R}}
\begin{document}
\title{Parity-violating hybridization in heavy Weyl semimetals}

\author{Po-Yao Chang}
\email{pychang@physics.rutgers.edu}
\affiliation{Center for Materials Theory, Rutgers University,
Piscataway, New Jersey, 08854, USA }

\author{Piers Coleman}
\affiliation{Center for Materials Theory, Rutgers University, Piscataway, New Jersey, 08854, USA }
\affiliation{Department of Physics, Royal Holloway, University of London, Egham, Surrey TW20 0EX, UK }

\begin{abstract}
We introduce a simple model to describe the formation of heavy Weyl semimetals
in non-centrosymmetric heavy fermion compounds under the influence of a
parity-mixing, onsite hybridization.
A key aspect of interaction-driven heavy Weyl semimetals is the
development of surface Kondo
breakdown, which is expected to give rise to 
a temperature-dependent re-configuration of the Fermi arcs and 
the Weyl cyclotron orbits which connect them via the 
chiral bulk states. Our theory predicts 
a strong temperature dependent transformation in the quantum
oscillations at low temperatures.  In addition to the effects of
surface Kondo breakdown, the renormalization effects in heavy Weyl
semimetals will appear in a variety of thermodynamic and transport
measurements.
%The enhancement of the non-saturate thermopower under magnetic field is discussed.
\end{abstract}

\maketitle
\section{Introduction}
%{\it Introduction---}
Heavy fermion materials are a tunable class of compounds in which
strong electron correlations give rise to a wealth of 
metallic, superconducting, magnetic and insulating phases.
A new aspect of these materials is the possibility of
topological behavior, epitomized
by the topological Kondo insulator (TKI) SmB$_6$\cite{Mott:1974ui,
Dzero2010, Dzero2012,
Alexandrov2013, Lu2013,Dzero2016}, in which 
a topologically non-trivial
entanglement between local moments and conduction electrons, gives
rise to Dirac surface states\cite{Jiang2013,Neupane2013,Xu2013,Frantzeskakis2013}.
An important second class of topological behavior occurs in the
presence of broken inversion or time-reversal symmetry, which
transforms the quantum critical point between normal and topological insulators
into a Weyl semimetal phase, with 
relativistic chiral fermions in the bulk and
Fermi arc states\cite{Murakami2007,Wan2011,Burkov2011} on the surface.
Various Weyl semimetallic phases 
 have been proposed and discovered in weakly interacting systems\cite{Xu2015,Weng2015,Huang2015_1}.
 Most Weyl semimetals are non-centrosymmetric crystals\cite{Murakami2007}.
 
 \begin{figure}[htbp]
\centering
\includegraphics[width=\columnwidth] {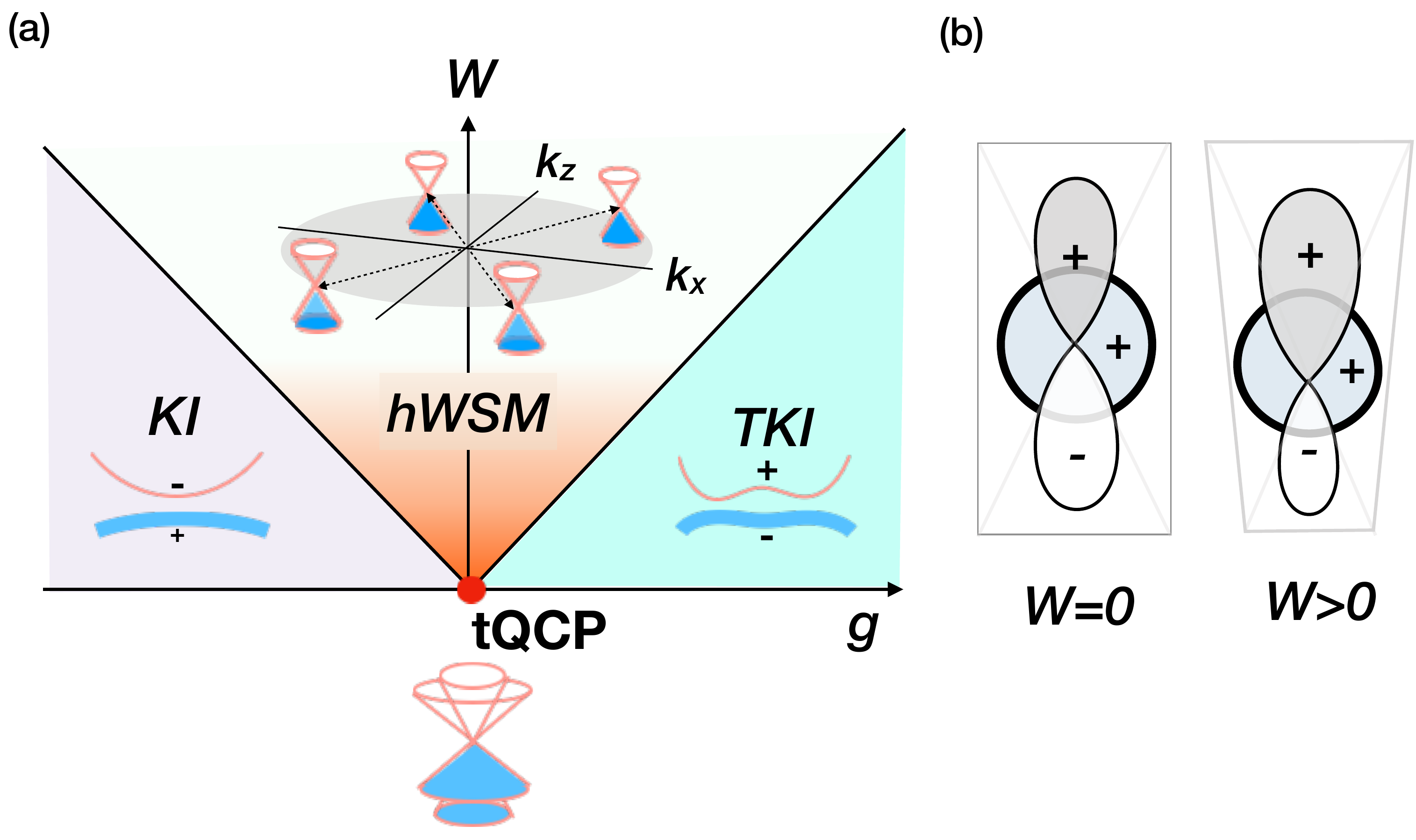}
\caption{(a) Topological Quantum
Critical Point (tQCP) at the nexus between normal and topological
Kondo insulators (KI/TKIs) and
heavy Weyl semimetals (hWSMs). $g$ and $W$ are the 
band tuning and inversion symmetry breaking parameters, respectively.
At the tQCP, charge neutrality pins the bulk Dirac cone to the Fermi energy
(occupied bands indicated by light blue).
Finite $W>0$  splits the Dirac point into
four symmetry-related Weyl points, pinned to the Fermi energy. 
(b) Breaking of inversion symmetry leads to a finite onsite
hybridization $W>0$.
}
\label{Fig_Phase_diagram}
\end{figure}

A preponderance of
noncentrosymmetric heavy fermion materials offers an exciting
opportunity to explore strongly interacting, or ``heavy Weyl semimetals'' (hWSMs)\cite{Xu2017,Lai2016}.  
Four
candidates have already come to light:  CeRu$_4$Sn$_6$\cite{Guritanu2013},
Ce$_3$Bi$_4$Pd$_3$\cite{Dzsaber2017},  CeRu$_4$Sb$_{12}$\cite{Okamura2011,Shankar2016}
and YbPtBi\cite{Guo2017}. 
Optical measurements on CeRu$_4$Sn$_6$\cite{Guritanu2013} and
transport measurements 
on CeRu$_4$Sb$_{12}$\cite{Okamura2011,Shankar2016} indicate
anisotropic semimetallic behavior. More remarkably, 
the recent observation of a giant $T^{3}$ component to the specific heat
of Ce$_3$Bi$_4$Pd$_3$\cite{Dzsaber2017} and YbPtBi\cite{Guo2017} reveals the presence of 
point-node excitations.
%precisely pinned to the Fermi energy.

 %In Refs. \cite{Wissgott2016,Xu2017}, density functional calculation including strong correlations
Recent density functional calculations\cite{Wissgott2016,Xu2017} 
confirm that heavy fermion systems are expected to host Weyl points
with surface Fermi arcs.  Lai et al.\cite{Lai2016}  have recently
proposed a tight-binding model for heavy Weyl semimetals\cite{Lai2016},
predicting that the density of states near the Weyl nodes is strongly
renormalized by the hybridization with f-electrons.
These works raise a number of open questions:

\begin{itemize}

\item what is the relationship between 
heavy Weyl semimetals and topological Kondo insulators?

\item beyond renormalization, what are the qualitative effects of strong interactions?

%\item how (if at all) are the Weyl cones pinned to the Fermi energy? 
\end{itemize}
In this letter, we propose 
a simple a two-band model  which links the emergence of heavy Weyl
semi-metals at the topological quantum critical point (tQCP)
between Kondo and topological Kondo insulators
to the development of a parity-breaking on-site
hybridization between $d$- and $f$-states in noncentrosymmetric Kondo lattices[Fig. \ref{Fig_Phase_diagram}(a)].

{One of the qualitative effects predicted by our model,
is the phenomenon of Kondo breakdown, whereby 
the loss of coordination of local moments at the surface
leads to a reduction of the surface Kondo temperature.
This phenomenon has been proposed as the origin of 
light surface quasiparticles observed in SmB$_6$\cite{Alexandrov2015}.
The analogous effect on 
the Fermi arcs causes them to 
reconfigure their geometry [Fig. \ref{Fig_2}] as a function of
temperature, giving rise to a strong temperature
dependence  in the inter-surface cyclotron orbits
\cite{Dai2016, Potter2014,Moll2016}.

\begin{figure}[htbp]
\centering
\includegraphics[width=\columnwidth] {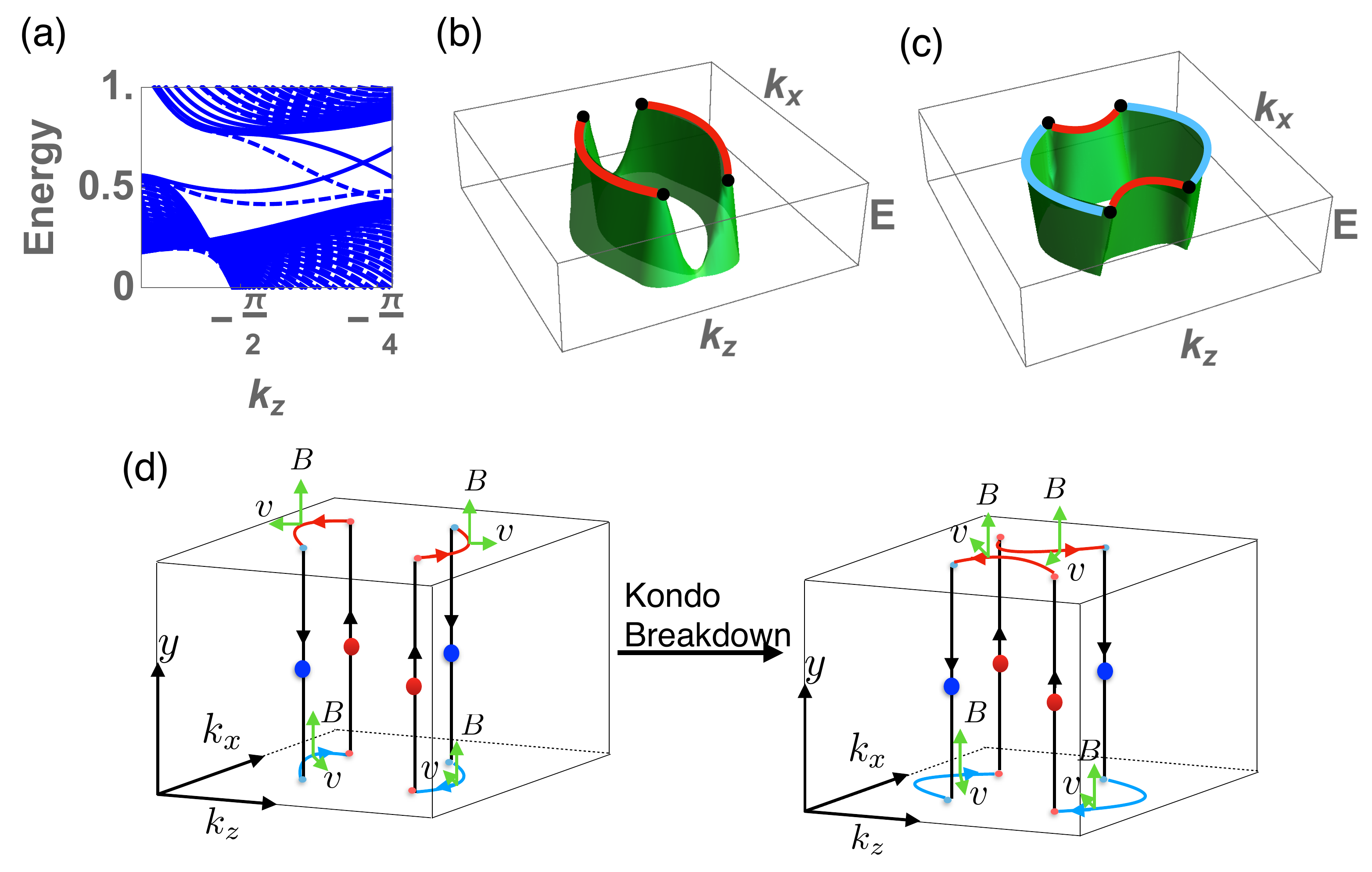}
\caption{(a) Kondo Breakdown in a heavy Weyl semimetal, contrasting 
the spectrum before (solid lines) and after (dashed lines)
surface Kondo breakdown 
as a function of $k_z$ at $k_x=0$. 
Surface spectrum (b) before
and (c) after surface Kondo breakdown:
red and blue lines indicate the Fermi arcs on the top and bottom surfaces respectively. Black dots indicate the projection of the 
Weyl nodes onto the surface.
Parameters were taken to be $(t_x,t_y,t_z,\mu,\alpha,V_x, V_y,V_z,W_2,b, \lambda)=(2,1,1,-6,-0.1,0.7,0.7,1.05,0.8,0.89,0.087)$  in Eq. (\ref{Ham}). 
(d) Schematic of the Weyl orbits, where arrows indicate the
quasiparticle trajectory. %and arcs on the top and bottom surfaces  are indicated by red and blue lines respectively.
%At low temperatures, without breakdown (left panel), there are two small Weyl
%orbits, but at intermediate temperatures 
% surface Kondo breakdown causes the 
%Fermi arcs to connect(right panel) via a single large Weyl orbit.
}
\label{Fig_2}
\end{figure}

Dzero et al. have emphasized that the spin-orbit 
driven topological behavior in  heavy fermion systems 
derives from the odd-parity hybridization between  $d$ ($\phi_{d}$)
and $f$-orbitals ($\phi_{f}$)\cite{Mott:1974ui,Dzero2010,Dzero2012} given by 
the Slater-Koster overlap integral
\begin{equation}\label{l}
\tilde{V}_{\alpha \beta } ({\bR}) = \int d^{3}x\phi^{*}_{d\alpha } ({\bf
x}-\bR) {\mathrm V}
(\bx) \phi_{f\beta } ({\bf x}), 
\end{equation}
where ${\mathrm V} (\bx )$ is the electronic potential.
Inversion symmetry in centrosymmetric crystals
fully eliminates the onsite hybridization 
between the opposite parity $d$ and $f$ states ($V_{\alpha\beta }
(0)=0$) [Fig. \ref{Fig_Phase_diagram}(b)], and
in momentum space, the residual intersite components of the 
hybridization  then acquire the odd-in time, odd-in momentum,
 relativistic form 
$V_{\alpha
\beta } (\bk )\sim \bk\cdot \vec{ \sigma }$ near the high symmetry
points.
The band-crossing permitted by this nodal hybridization 
drives the formation of topological Kondo insulators. 
{On the other hand, in 
%A first observation of this paper
%is that  topological Kondo insulators are nascent Weyl semi-metals. 
non-centrosymmetric crystals, 
the asymmetric potential ${\rm
V}(\bx)\neq {\mathrm V} (-\bx)$ distorts the $f$ and $d$ 
orbitals and eliminates parity conservation, 
giving rise to a finite onsite d-f hybridization $W_{\alpha \beta }
=\tilde{V}_{\alpha \beta
} (\bR=0) $[Fig.  \ref{Fig_Phase_diagram}(b)]. 
Under the influence of this perturbation, 
topological Kondo insulators become heavy Weyl semimetals
as shown in Fig.  \ref{Fig_Phase_diagram}(a).
}
A minimal model for the hybridization that captures 
these features in a two-band model is obtained by generalizing the nearest-neighbor model
introduced by 
Alexandrov, Coleman, and Erten\cite{Alexandrov2015} (ACE) to include an
additional onsite term as follows:
\begin{eqnarray}\label{hybrid}
\tilde{V}({\bf R})_{\alpha\beta}&= & \left\{ 
\begin{array}{lc}\displaystyle
- {i} \vec{ v}_{{\bf R}}\cdot {\boldsymbol{\sigma }}_{\alpha \beta }
,  &\quad {{\bf R}\in \hbox{n.n.}}  \\
 {w}_{0}+i\vec{{w}}\cdot {\boldsymbol{\sigma}} &\quad {{\bf R}=0}.
\end{array}
\right.
\end{eqnarray}
where the vector $\vec{v}_{\bf R}= (v_{1}R_{1},v_{2}R_{2},v_{3}R_{3})
% {\rm Diag}(v_{1}, v_{2},v_{3})\cdot\mathbf{R}
$
describes the strength of the nearest neighbor hybridization while
$w_{0}$ and $\vec{w}$ describe the inversion-symmetry breaking onsite
hybridization terms, in a time-reversal invariant form.

\section{Model}
%{\it Model---}
We use a non-centrosymmetric modification of the ACE model
\cite{Alexandrov2015} 
\begin{equation}\label{}
H=\sum_{i,j,\sigma,\sigma'} \Psi_{i\sigma}^\dagger \mathcal{H}_{ij,\sigma\sigma'}\Psi_{j\sigma'}+U\sum_i n_{if \uparrow} n_{if \downarrow},
\end{equation}
where
%\begin{widetext}
\begin{align}
%H=\sum_{i,j,\sigma, \sigma'} \left(\begin{array}{cc}c^\dagger_{i,\sigma}, & f^\dagger_{i,\sigma}\end{array}\right)
\mathcal{H}_{ij,\sigma\sigma'}=\left(\begin{array}{cc} (
-t^c_{i,j}-\mu^c\delta_{ij})\delta_{\sigma \sigma '} &
\tilde{V}_{\sigma \sigma'}({\bf R}_i-{\bf R}_j) \\ \tilde{V}_{\sigma
\sigma'}({\bf R}_i-{\bf R}_j) & (
-t^f_{i,j}-\mu^f\delta_{ij})\delta_{\sigma \sigma
'}\end{array}\right).
%\left(\begin{array}{c}c_{j\sigma'} \\f_{j\sigma'}\end{array}\right)
%+U\sum_i n_{if \uparrow} n_{if \downarrow}.
\end{align}
Here $\Psi_{i\sigma}^\dagger=(c^\dagger_{i\sigma}$, $f^\dagger_{i\sigma})$ with $c^\dagger_{i\sigma}$ and $f^\dagger_{i\sigma}$ are the creation operators
for conduction and f-electrons.
%\end{widetext}
%where $c^\dagger_{i\sigma}$ and $f^\dagger_{i\sigma}$ are the creation operators
%for conduction and f-electrons, respectively.
$t^{c/f}_{ij}$ is the hopping amplitude and $\mu^{c/f}$ is the
chemical potential for $c/f$ electrons.
$U$ is the onsite Coulomb repulsion between f-electrons. 
%The form of the hybridization 
%$\tilde{V}_{\sigma \sigma'}({\bf R}_i-{\bf R}_j)$ is given in 
%(\ref{hybrid}).

In the large $U$ limit, a slave-boson approach leads to the mean-field Hamiltonian\cite{Coleman1987}, 
$H=\sum_{\bf k}\Psi^\dagger({\bf k})\mathcal{H}({\bf k})\Psi({\bf k})+\mathcal{N}_s\lambda(|b|^2-Q)$ with
\begin{align}
\mathcal{H}({\bf k})=&\left(\begin{array}{cc}\epsilon_c({\bf k})-\mu & \sum_j V_j \sigma_j \sin k_j \\
\sum_j V_j \sigma_j \sin k_j  & \epsilon_f({\bf k})+\lambda\end{array}\right)  \notag \\
&+ \left(\begin{array}{cc} 0 & W_0 +i \vec{W}\cdot \vec{\sigma }
 \\
W_0 -i \vec{W}\cdot \vec{\sigma } & 0 \end{array}\right).
\label{Ham}
\end{align}
$V_i=v_ib$ and $W_{i}={w}_ib$ are the renormalized hybridization terms, $b$ is the slave boson projection amplitude.
The f-hopping amplitude becomes $\tilde{t}^f=b^2 t^f$
The dispersion of the conduction electrons is taken as $\epsilon_c({\bf k})=-2\sum_i t_i \cos k_i $
and $\epsilon_f({\bf k})=\alpha \epsilon_c({\bf k})$ for simplicity.
The constraint field $\lambda$ imposes the mean-field constraint $Q=n_f+b^2$ with
$Q$ being the local conserved charge associated with the slave boson approach in the infinite $U$ limit,
and is taken to be $Q=1$. $\mathcal{N}_s$ is the total number of sites.
We solve the slave-boson mean-field Hamiltonian self-consistently [see Appendix \ref{App:A}].

The spectrum of the Hamiltonian [Eq. (\ref{Ham})] %in the presence of the inversion symmetry breaking terms 
is
\begin{align}
E({\bf k})=h_0 \pm \sqrt{h_1^2+W_\mu^2 +\vec{V}_{\bk }^{2}
\pm 2\sqrt{
 W_\mu^2 \  \vec{V}_{\bk }^{2}- (\vec{W}\cdot\vec V_{\bk })^{2}  }
},
\label{Eq:spec}
\end{align}
where $h_{0/1}=\frac{1}{2}[\epsilon_{f}({\bf k})+\lambda\pm (\epsilon_c({\bf
k})-\mu)]$, $W_\mu^2=
W_{0}^{2}+\vec{W}^{2}$ and $\vec{V}_{\bk }= (V_{1}\sin k_{1},
V_{2}\sin k_{2}, V_{3}\sin k_3)$.

The energy spectrum has line or point nodes determined by the
intersections between three surfaces:
${\cal  S}_{I}$ where
$h_1 =0$, ${\cal S}_{II}$, where
$(W_\mu^2 -\vec{V}_{\bk }^{2})^2=0$ and  ${\cal S}_{III}$ where
$\vec{W}\cdot\vec{V}_{\bk }=0$. 
When there is no common intersection between these
surfaces, the ground-state remains a fully gapped insulator.
However, once $W$ exceeds a critical value, a semi-metallic state
develops.  There are two cases:
\begin{itemize}
\item Line-node semimetal ($\vec{W}=0$) for which the constraint
${\cal S}_{III}$ is trivial. Since ${\cal S}_{I}$ and ${\cal S}_{II}$ are
spheroids that share the same center,  they intersect to form two rings 
$\{r_{1},r_{2} \} = {\cal S}_{I}\cap {\cal S}_{II}$
of gapless excitations [Fig. \ref{Fig:S123}(a)] [see Appendix \ref{App:E} for the detailed discussion].

\item Weyl semimetal ($\vec{W} \neq 0$). Here ${\cal S}_{III}$ is the plane normal to $\vec{W}$, intersecting 
with rings $\{r_{1},r_{2}\}$ at four Weyl points [Fig. \ref{Fig:S123}(b)].

\end{itemize}

\begin{figure}[htbp]
\centering
\includegraphics[height=2.3 cm] {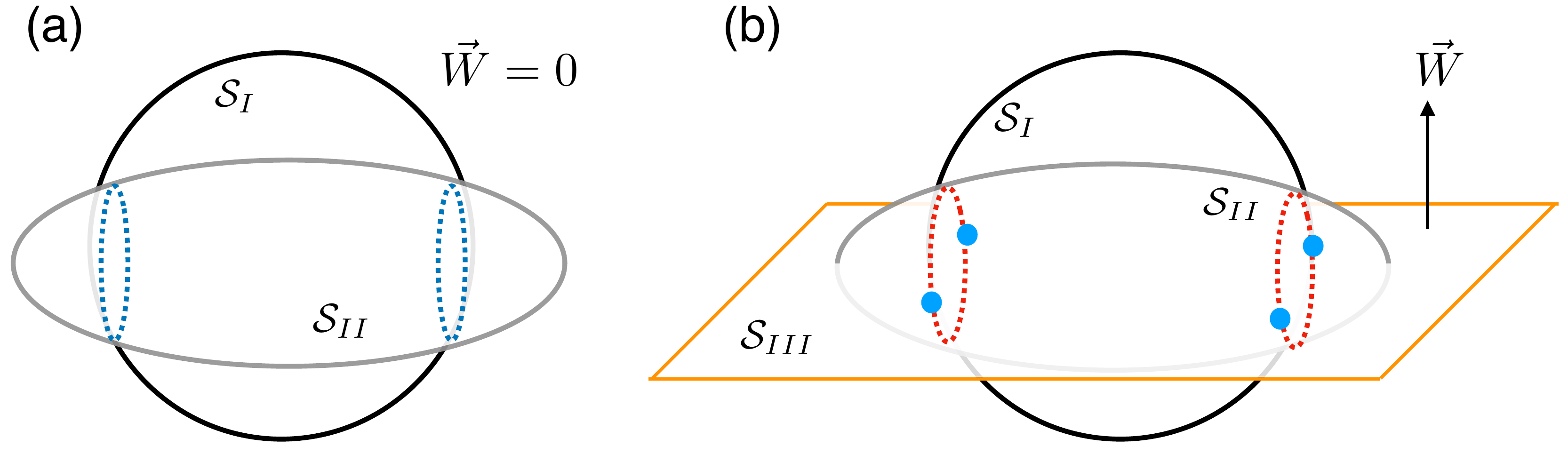}
\caption{
(a) A line-node semimetal: when $\vec{W}=0$, the line-node solution from
Eq. (\ref{Eq:spec}) is determined by the intersections
$\{r_1,r_2\}={\cal S}_{I}\cap {\cal S}_{II}$ indic
ated by
blue  lines.
(b) A Weyl semimetal: When $\vec{W}\neq 0$, 
point-node solutions of Eq. (\ref{Eq:spec}), indicated by blue dots,  develop at the
intersections of plane ${\cal S}_{III}$ normal to $\vec{W}$ with
the line-nodes, ${\cal S}_{I}\cap {\cal S}_{II}\cap {\cal S}_{III}$.
}
\label{Fig:S123}
\end{figure}

Time reversal and reflection 
symmetries play an important role in Weyl semimetals. Our model
preserves time-reversal symmetry, $\mathcal{T}^{-1} \mathcal{H}({\bf k})
\mathcal{T}= \mathcal{H}(-{\bf k})$,
where $\mathcal{T}=i\sigma_2 \mathcal{K}$ and 
$\mathcal{K}$ is the conjugation
operator. In the absence of $W_0$, providing
the inversion-symmetry breaking vector $\vec{W}=W\hat l$ points along a crystal axis $\hat
l$, then the model also retains reflection 
symmetries in the two planes with normals $\hat j\neq  \hat l$ 
perpendicular to $\hat l$.
For our model, the reflection operator is
$\mathcal{R}_j=\sigma_j \tau_3$ and
$\mathcal{R}^{-1}_j \mathcal{H}({\bf k}) \mathcal{R}_j =
\mathcal{H}({{\bf k_{R}}})$, where ${\bf k_{R}}$ is
the wavevector reflected in the plane perpendicular to $\hat j$ and
${\boldsymbol\tau}$ are the Pauli matrices in 
 $c,f$  space. 
%{\clr We should note that $W_0$ term describes a bound state with high spin $ \langle c^\dagger_{\uparrow} f_{\uparrow}+c^\dagger_{\downarrow} f_{\downarrow} \rangle$,
%which is energetic and we will set it to be zero in the following discussion.
%Furthermore,
Suppose for example $\vec{W}= W_2 \hat  y$, 
then the energy spectrum has four Weyl points located
in the  $k_y=0$ plane, each related to one-another by time reversal
and reflection symmetries $\mathcal{R}_x$ and $\mathcal{R}_z$.

The effective  Hamiltonian near the Weyl points is 
obtained from Eq. (\ref{Ham}) by projecting it onto the
eigenvectors of the two central bands [see Appendix \ref{App:B}].  
For $\vec{W}=W_{2}\hat y$, we have four Weyl points $\vec{k}_0$,
$R_x\vec{k}_0=(-k_{x0},0,k_{z0})$, $R_z\vec{k}_0=(k_{x0},0,-k_{z0})$, and $T \vec{k}_0=- \vec{k}_0$ related by
reflection and time-reversal symmetries, respectively.
The effective Hamiltonian can be expressed in a general form
 \begin{align}
 \mathcal{H}_{\rm eff}(\vec{k}_0,\delta {\bf k})=&[A({\vec{k}_0})]_{i \alpha} \delta k_i \sigma_\alpha.% + [B_{\vec{k}_{0}}]_{i}\delta  k_{i}\  \mathbb{I}_{2 \times 2}.
 \label{Eq:Weyl}
 \end{align}
with implied summation on $i\in [x,y,z]$ and $\alpha \in [0,1,2,3]$ with 
$\sigma_0= \mathbb{I}_{2 \times 2}$. Here
$[A({\vec{k}_0})]_{i\alpha}$ is a three by four matrix
defined at each Weyl point $\vec{k}_0$, each proportional
to the hybridization amplitudes $V_i$ [see Appendix \ref{App:B}].
These four effective Hamiltonians 
are related by reflection and time-reversal symmetries 
($R_{x(z)}:  \mathcal{H}_{\rm eff}( {\vec{k}_0},\delta {\bf k})  \to  \mathcal{H}_{\rm eff}( {R_{x(z)}\vec{k}_0},R_{x(z)}\delta {\bf k})$ and
$T:  \mathcal{H}_{\rm eff}( {\vec{k}_0},\delta {\bf k})  \to
\mathcal{H}_{\rm eff} ({-\vec{k}_0},-\delta {\bf k})$) which 
constrains the  four Weyl points to lie at the same energy.
%This symmetry, combined with the 
%charge-neutrality condition $n_{ci}+n_{fi}=2$  
%enforces the Fermi energy to lie at the gap-closing point, pinning the 
%four Weyl points to the Fermi energy. 
%This pinning effect offers an explanation of the pristine $T^3$ behavior of the 
%heat capacity observed
%in the candidate hWSM material Ce$_3$Bi$_4$Pd$_3$\cite{Dzsaber2017}.

%In principle, inhomogeneities can cause he chemical potential to develop spatial fluctuations,
%leading to puddles of electron or hole doped semi-metal. 
%However, in heavy fermion materials, the $f$-electrons are highly
%incompressible $d \mu/d n_f \sim 0$ and the Weyl cones float on the
%Fermi energy to maintain this constraint, with the result that the
%formation of puddles is highly suppressed. 

\section{Surface Kondo breakdown}
%{\it Surface Kondo breakdown---}
We now examine the effect of ``Kondo breakdown'' on the Fermi arcs.
The topological charges $C=\pm 1$ [see Appendix \ref{App:C}] associated with the Weyl points
give rise to the formation of Fermi arcs 
%We denote it is the topological charge of the Weyl points.
which link the projections of oppositely charged Weyl points 
onto the surface Brillouin zone (BZ).
The analytic form of the localized Fermi arcs can be derived from the effective Hamiltonian [see Appendix \ref{App:D}].
In the presence of interactions, 
the reduction in co-ordination number of the f-elections at the
surface suppresses the surface Kondo temperature $T_{K}^{*}$ below that of the
bulk, 
$T^s_K< T_K$.
In the intermediate temperature regime $T^s_K<T< T_K$, the bulk is
topological but the conduction electrons at 
the surface are liberated from the local moments, 
leading to surface Kondo breakdown.
The surface Kondo breakdown scenario has been confirmed in inhomogeneous mean-field approach\cite{Alexandrov2015}
and dynamical mean field calculations \cite{Peters2016}. 
To model this effect, we suppress the slave boson amplitude $b$ to zero
on the surface layer of hWSMs and recompute the Fermi arcs. 

The effect of Kondo breakdown on the 
surface spectra for a $(010)$ slab geometry 
is shown in Figs. \ref{Fig_2}(a)-(c):
%The red lines indicate the crossings of the right and left chiral
%modes on opposite surfaces. 
we see that the intersections between two surface chiral modes sink
beneath the Fermi sea.  This effect causes the right and left chiral
modes to bulge outwards, leading to a differential reconfiguration of
the Fermi arcs on opposite surfaces as shown in Fig. \ref{Fig_2}(c).
 In fact, the detailed configuration of the Fermi arcs will in
general depend on the microscopic parameters of the Hamiltonian. For
example, in CeRu$_4$Sn$_6$ \cite{Xu2017}, the nonequivalent cleavage
surfaces are found to give rise to different configurations of Fermi
arcs.  This indicates that the Fermi arcs are sensitive to the surface
morphologies and chemical potential.  The configuration of the Fermi
arcs will also be sensitive to the surface hybridization.  Thus the
surface Kondo breakdown introduces the reconfiguration of the Fermi
arcs. The configuration of the Fermi arcs is a global property of the
system, dependent on both bulk and surface properties.  In particular,
the configuration depends on the locations of the projected Weyl
points on the surface BZ and the detailed dispersions of the surface
spectrum.  On the other hand, the topology of each Weyl point is a
local property, with a generic form as described by
Eq. (\ref{Eq:Weyl}).  The finite topological charge $C=\pm 1$ of the
Weyl point, ensures the formation of Fermi arcs which link with the
projections of oppositely charged Weyl points on the surface
BZ. However, this
local property does not constrain the way the pairs of
oppositely charged Weyl points are linked.  

%\begin{figure}[htbp]
%\centering
%\includegraphics[height=3.3 cm] {arc_orbit.pdf}
%\caption{A schematic plot for the Weyl orbit. The black arrows indicate the directions of the quasiparitcle trajectory. The Lorentz force $\dot{\bf k}=q {\bf v}_{\rm S} \times {\bf B}$ acts on the quasiparitcle
%on top and bottom surfaces, where ${\bf v}_{\rm S}$ is the surface velocity, $q=-e$ is the charge of the quasiparitcle, and ${\bf B}$ is
%the magnetic field. In the bulk, there are chiral Landau levels $\epsilon(k_y)_{\pm}=\pm v_{\rm B} k_y$ for the $\pm 1$ Berry curvature of Weyl points respectively
%and $v_{\rm B}$ is the bulk velocity.  
%The Fermi arcs of the top and bottom surfaces are connected by the bulk channel from the chiral Landau level and form close loops (Weyl orbits).
%Left panel: before surface Kondo breakdown, there are four Weyl points in the bulk and two Weyl orbits.
%Right panel: after surface Kondo breakdown, the Fermi arc surface states get reshaped and form a large Weyl orbit.}
%\label{Fig_Weyl_Orbits}
%\end{figure}

%\subsection{Experimental consequences: quantum oscillations on Weyl orbit and renormalization effects}
%{\it Experimental consequences: quantum oscillations on Weyl orbit
%and renormalization effects---}
The reconfiguration of the Fermi arcs will have various distinct
signatures in both angle-resolved photoemission spectroscopy and quantum oscillation
measurements. 
In a field, quasiparticles on the surface 
move under the influence of the Lorentz force $\dot{\bf k}=-e {\bf
v}_{\rm S} \times {\bf B}$, where ${\bf v}_{\rm S}$ is their velocity,
processing from one projected Weyl point to another. 
When they reach a Weyl
point, the gapless bulk chiral Landau level
%$\epsilon(k_y)_{\pm}=\pm
%v_{\rm B} k_y$ ($\pm$ denotes the sign of topological charge) 
provides a 
transport channel to coherently move the quasiparticles between 
surfaces, giving rise to closed inter-plane Weyl orbits, 
\cite{Dai2016, Potter2014} as shown in
Fig. \ref{Fig_2}(d).  Quantization of the Weyl orbital
motion leads to discrete energy levels 
%\begin{align}
$E_n=\frac{\pi h (n+\gamma) v_{\rm B }}{L+\beta k_0 /(e B)}$,
%\end{align}
where $k_0$ is the length of the Fermi arcs, $\mu$ is the chemical potential, $L$ is the thickness of the sample,
$\gamma$ is a constant, and $\beta = v_{\rm B} /v_{\rm S}$ with $ v_{\rm B}$ being the bulk velocity.
Such Landau levels have  been observed in 
Shubnikov-de Haas oscillations in Cd$_3$As$_2$, 
a weakly interacting Dirac semimetal which is the crystal-symmetry-protection analogy of a Weyl semimetal\cite{Moll2016}.

One of the most dramatic consequences of the differential
reconfiguration is the merger 
of two small orbits into a single large orbit as shown in
Fig. \ref{Fig_2}(d), and the effect that will modify
the quantum oscillations. 
Suppose the chemical potential is fixed to be $\mu$ and vary the magnetic field $B$,
the $n$th energy level crosses the $\mu$ with the condition 
\begin{align}
\frac{1}{B_n}=
\begin{cases}
\frac{e}{ \beta k_0} \left(\frac{v_{\rm B} \pi h}{\mu} (n+ \gamma)-L \right),& T<T_K^s<T_K\\
\frac{e}{ \beta'(k_0'+k_1)} \left(\frac{ v_{\rm B}  \pi h}{\mu} (n+ \gamma)-2L \right),& T_K^s<T<T_K,
\end{cases}
\end{align}
where $k_0'$ and  $k_{1}$ are the arc-lengths on the 
bottom and top surfaces respectively [see right panel in Fig. \ref{Fig_2}(d)],
while $\beta'=v_{\rm B}/v'_{\rm S}$ with $v'_{\rm S}$ being the surface velocity of quasiparticles with surface Kondo breakdown.

During the transition of surface Kondo breakdown,
the spacing of the density of states as a function of $1/B$ has a dramatic change, %going from
$\omega_{B^{-1}} =\frac{e \pi h v_{\rm B}}{ \beta \mu k_0} \to
\frac{e \pi h v_{\rm B}}{ \beta' \mu (k'_0+k_1)}$.
The magnetic field $1/B$ threshold of observing this oscillation also changes from $e L/ \beta k_0 \to 2e L/ \beta' (k'_0+k_1)$. 
The differential reconfiguration of the Fermi arc states can be detected by measuring 
the change of oscillation frequency and a threshold of the magnetic field in Shubnikov-de Haas oscillations.

\section{Renormalization effects}
%{\it Renormalization effect---}
%In addition to the effect of surface Kondo breakdown,
The renormalized velocity of the Weyl semimetals described in
Eq. (\ref{Eq:Weyl}) is proportional to the hybridization amplitude
$V_i \propto \sqrt{T_K D}$ where $T_K$ is the Kondo temperature and
$D$ is the band width of the conduction electrons\cite{Colemanbook}.
This ``square-root'' renormalization effect 
is weaker than that seen in heavy fermion metals, due to the hybridization
origin of the nodes. 
From Ref. [\onlinecite{Xu2017}], the velocity of Weyl fermion in CeRu$_4$Sn$_6$
is $v^*\sim 0.2$ eV\AA.  For the weakly interacting Weyl semimetals
such as TaAs\cite{Lv2015} and TaP\cite{Xu2015}, the velocity of Weyl
fermion $v\sim 2-3$ eV\AA. The renormalization effect in hWSMs is
about a factor of ten.

Many of the thermodynamic and transport properties in hWSMs are
affected by this quasiparticle renormalization effect.
One of the most dramatic effects, is the renormalization of the cubic
specific heat. A large 
$T^{3}$ specific heat 
has been reported in the candidate hWSM materials
Ce$_3$Bi$_4$Pd$_3$\cite{Dzsaber2017} and YbPtBi\cite{Guo2017}.
%At first sight, the $T^3$ behavior in specific heat is rather 
%surprising because it strongly suggests that Weyl points are
%pinned to the Fermi energy.   There are two factors at work here that
%connect with our model. 
%First, crystalline symmetries act together with
%time-reversal symmetry to constrain the four Weyl points to lie at the
%same energy. Charge neutrality then assures that the four Weyl points
%lie at the Fermi energy. 
%However, equally important is the incompressible
%nature of the f-electron fluid.  In a conventional Weyl system,
%inhomogeneities in 
%the chemical potential occur against a 
%rigid band-structure, locally elevating or immersing the Weyl cones
%in the Fermi sea to create 
%a finite density of states 
%proportional to the variance in the chemical potential
%$\delta \rho \propto \langle \delta \mu^{2} (\bx
%)\rangle $.  This would normally create a finite linear specific heat
%masking the enhanced $T^{3}$ specific heat. 
%However, in a Kondo lattice,
%the incompressible nature of the f-fluid (imposed by the constraint
%$n_{f}+b^{2}=1$) guarantees that the f-level floats on the Fermi
%energy to maintain a constant f-valence, 
%so that  $\delta \lambda (\bx)= \delta  \mu (\bx )$ which
%causes the Weyl cones to {\sl float} on the Fermi energy
%[Fig. \ref{Fig:Floating}].
As pointed out by Lai et al.\cite{Lai2016} 
this significant 
enhancement of specific heat likely derives 
from the  cubic dependence on renormalized velocity 
 $C_v = \frac{\partial}{\partial T} \int \epsilon f(\epsilon) g(\epsilon) d\epsilon \propto (T/v^*)^3$ 
with $g(\epsilon)=\frac{\epsilon^2}{2 \pi^2 v^{*3}}$ being the density of states.
%\begin{figure}[htbp]
%\centering
%\includegraphics[height=1.35 cm] {Floating.pdf}
%\caption{
%(a) Weyl points can be locally 
%elevated or immersed at the Fermi surface for conventional Weyl semimetals.
%(b) Weyl points float on the Fermi surface for hWSMs. Blue regions indicate the occupied bands.
%}
%\label{Fig:Floating}
%\end{figure}
In addition to the specific heat, 
an enhancement of the high-field thermopower\cite{Skinner2017}
is also expected.
The high-field thermoelectric properties of 
the Weyl/Dirac semimetals contrast dramatically 
with those of doped semiconductors, with a 
thermopower that grows linearly, without saturation, in a 
the magnetic field,
$\alpha:=\Delta V/ \Delta T  \propto B T /v^*$, where $\Delta V$ and
$\Delta T$ are the voltage and temperature difference,  respectively. 
The non-saturating behavior leads 
to a large thermopower which has been observed in weakly interacting Dirac semimetal Pb$_{1-x}$Sn$_x$Se\cite{Liang2013}.
The high-field thermopower is thus enhanced by the mass
renormalization in hWSMs.

%Interestingly, the electrical and thermal conductivity is not renormalized,
%$\sigma_{xx} = \frac{e^2}{3}v^{*2} \tau g(\epsilon)|_{\epsilon=\mu} \propto {\rm const.}$ and
%$\kappa_{xx} \propto T \sigma_{xx} \propto {\rm const.}$.
%And finally, the magneto-conductivity depends on the square of the renormalized velocity
%$\sigma_{\rm CME} = \frac{e^4 B^2}{4 \pi^2 \hbar c^2 \mu^2} v^{*3} \tau
%\propto  v^{*2} B^2$\cite{Son2013}. }
%The magneto conductance 
%leads to negative magnetoresistance observed in TaAs, NbAs, and
%NbP\cite{Huang2015,Li2017}: in hWSMs, we expect these effects to be
%heavily suppressed by their mass enhancement. 
 
\section{Conclusion and discussion}
%{\it Conclusion and discussion---}
We have proposed a hybridization-driven model for heavy
Weyl semimetals, arguing
that the onsite hybridization between $f$ and $d$ orbitals in non-centrosymmetric
crystals drives topological Kondo insulators into hWSMs. One of the
effects of the strong interactions is 
surface Kondo breakdown, which 
leads to a reconfiguration of Fermi arcs on both surfaces that should
appear in quantum oscillations, while the renormalization of velocity in hWSMs affects thermodynamic and transport properties.
%We have argued that the charge-neutrality conduction, combined 
%with symmetries and the incompressible nature of the f-electron fluid
%pins the Weyl point to the local chemical potential. 
%This  effect accounts for the $T^3$ behavior of the heat capacity observed in Ce$_3$Bi$_4$Pd$_3$ and YbPtBi.

There are a number of interesting new directions for research into
hWSMs that deserve mention.  One aspect that deserves exploration
is the influence of non-symmorphic space group symmetries.  According to
topological band theory\cite{Bradlyn2016}, such symmetries can lead to
nodal points with much higher multiplicities, giving rise to a cluster
of nested Dirac cones. A particularly interesting case is the
candidate hWSM Ce$_3$Bi$_4$Pd$_3$, the space group  No. $220$ ($I\bar{4}3d$)  is expected to produce
an eight-fold degenerate double Dirac point.
These nodal lines are expected to give rise to ``drumhead surface states''\cite{Matsuura2013, Burkov2011, Bian2016} [see Appendix \ref{App:E}], which can potentially cause
charge/spin density wave and superconducting instabilities.
A second interesting direction, 
is the possible use of molecular beam
epitaxy (MBE) techniques 
\cite{Ishii2016}, which open up the possibility of  artificially
engineered hWSMs where 
%{\clr As emphasized in the paper, the pinning effect in topological heavy fermion systems 
%provides a nature playground for studying the topological quantum critical point.
%Using MBE techniques, 
tuning 
the degree of inversion symmetry breaking can be 
used to explore the vicinity, and possible
instabilities of the topological quantum critical point\cite{Yang2014, Isobe2016}.}

\begin{acknowledgments}
The authors would like to thank James Analytis, 
Elio K\"onig, Silke Paschen  and Yuanfeng Xu for valuable discussions. 
This work was supported by the Rutgers Center for Materials Theory group postdoc grant
(Po-Yao Chang) %, US National Science Foundation grant grant DMR-1309929 (Po-Yao Chang) 
and US Department of Energy grant DE-FG02-99ER45790 (Piers Coleman).
\end{acknowledgments}

\appendix

\section{Self-consistent slave boson mean-field solutions}
\label{App:A}

In the large $U$ limit, a slave-boson approach leads to the mean-field Hamiltonian\cite{Coleman1987}, 
$H=\sum_{\bf k}\Psi^\dagger({\bf k})\mathcal{H}({\bf k})\Psi({\bf k})+\mathcal{N}_s\lambda(|b|^2-Q)$ with
\begin{align}
\mathcal{H}({\bf k})=&\left(\begin{array}{cc}\epsilon_c({\bf k})-\mu & \sum_j V_j \sigma_j \sin k_j \\
\sum_j V_j \sigma_j \sin k_j  & \epsilon_f({\bf k})+\lambda\end{array}\right)  \notag \\
&+ \left(\begin{array}{cc} 0 & W_0 +i \vec{W}\cdot \vec{\sigma }
 \\
W_0 -i \vec{W}\cdot \vec{\sigma } & 0 \end{array}\right).
\end{align}
$V_i=v_ib$ and $W_{i}={w}_ib$ are the renormalized hybridization terms, $b$ is the slave boson projection amplitude.
The f-hopping amplitude becomes $\tilde{t}^f=b^2 t^f$
The dispersion of the conduction electrons is taken as $\epsilon_c({\bf k})=-2\sum_i t_i \cos k_i $
and $\epsilon_f({\bf k})=\alpha \epsilon_c({\bf k})$ for simplicity.
The constraint field $\lambda$ imposes the mean-field constraint $Q=n_f+b^2$ with
$Q$ being the local conserved charge associated with the slave boson approach in the infinite $U$ limit,
and is taken to be $Q=1$. $\mathcal{N}_s$ is the total number of sites.

The saddle point equations can be obtained from
$\frac{\delta \langle H \rangle}{\delta b}=0$ and $\frac{\delta \langle H \rangle}{\delta \lambda}=0$.
\begin{align}
&\frac{1}{\mathcal{N}_s }\sum_{{\bf k}, \sigma} \langle f^\dagger_{{\bf k},\sigma } f_{{\bf k},\sigma } \rangle+b^2=1 \\
&\frac{1}{2 \mathcal{N}_s }\sum_{{\bf k}, \sigma,i,j, \alpha, \beta} \{ (v_i \sin k_{i} [\sigma_i]_{\alpha \beta}+w_0\delta_{\alpha \beta} +i w_j[\sigma_j]_{\alpha \beta})\\
& \langle c^\dagger_{{\bf k},\alpha} f_{{\bf k}, \beta} \rangle
+h.c. \}+b(\lambda - \frac{1}{\mathcal{N}_s }\sum_{{\bf k}, \sigma} \langle  \tilde{ \epsilon}_{f}({\bf k}) f^\dagger_{{\bf k},\sigma } f_{{\bf k},\sigma } \rangle)=0,
\end{align}
where $\tilde{ \epsilon}_{f}({\bf k}) =\frac{1}{b^2} { \epsilon}_{f}({\bf k})$ is the bare spectrum of the f-electron.

In the paper we consider the case of non-zero 
on-site hybridization $w_2 \neq 0$. For the specific calculations
carried out in the paper, we have chosen the  bare parameter values to be 
$(v_x,v_y,v_z,w_2, \alpha, t_x, t_y, t_z,
\mu)=(0.786,0.786,1.179,0.89,-0.126,2,1,1,-6)$, leading to a 
self-consistently determined slave boson amplitude and the constraint field
with values $(b,\lambda)=(0.89, 0.087)$.

\section{Effective two dimensional Hamiltonian near the Weyl points}
\label{App:B}

Now we analyze the effective Hamiltonian around the Weyl points.
We consider the inversion breaking hybridization $W_2 \neq 0$ such that the Weyl points are located at $k_y=0$ plane.
The locations of the Weyl points in momentum space satisfy 
\begin{align}
&-2t_x\cos k_{x0} -2 t_z \cos k_{z0}-(\mu+2 t_y)=0, \notag\\
&W_2^2=V_1^2\sin^2 k_{x0} +V_3^2 \sin^2 k_{z0}.
\end{align}
We can expand the Hamiltonian around the Weyl points up to linear terms in ${\bf k}$.
\begin{align}
\mathcal{H}(\delta{\bf k} )\sim \mathcal{H}_0+\mathcal{H}_1(\delta{\bf k} ),
\end{align}
where
\begin{align}
\mathcal{H}_0=
[V_1(\sin k_{x0})\sigma_1+V_3(\sin k_{z0})\sigma_3]\tau_1+W_2\sigma_2\tau_2,
\end{align}
and
\begin{align}
\mathcal{H}_1(\delta{\bf k} )=& \frac{1}{2} (1+\alpha) [2t_x \sin k_{x0} \delta k_x+2t_z \sin k_{z0} \delta k_z]\tau_3 \notag\\
&+[V_1(\cos k_{x0} \delta k_x)\sigma_1+V_2 \delta k_y \notag\\
&+V_3(\cos k_{z0}\delta k_z)\sigma_3]\tau_1.
\end{align}
Here we have dropped terms proportional to the identity matrix which
only shift the spectrum without changing the band topology.  To obtain
the effective two-dimensional Hamiltonian in the vicinity of the Weyl points
we first find two eigenvectors of $\mathcal{H}_0$ with zero energy,
\begin{align}
|v_{1}\rangle=&\frac{1}{\sqrt{2(V^2_{1,k_{x0}}+V^2_{3,k_{z0}})+2V_{1,k_{x0}}\sqrt{V^2_{1,k_{x0}}+V^2_{3,k_{z0}}} }} \notag\\
&\left(\begin{array}{c}
V_{1,k_{x0}}+\sqrt{V^2_{1,k_{x0}}+V^2_{3,k_{z0}}} \\
0 \\
-V_{3,k_{z0}} \\
0
\end{array}\right),  \notag\\
|v_{2}\rangle=&\frac{1}{\sqrt{2(V^2_{1,k_{x0}}+V^2_{3,k_{z0}})+2V_{1,k_{x0}}\sqrt{V^2_{1,k_{x0}}+V^2_{3,k_{z0}}} }} \notag\\
&\left(\begin{array}{c}
0 \\
V_{3,k_{z0}} \\
0 \\
V_{1,k_{x0}}+\sqrt{V^2_{1,k_{x0}}+V^2_{3,k_{z0}}}
\end{array}\right).
\end{align}
where 
$V_{1(3),k_{x/z0}}=V_{1(3)}\sin k_{x(z)0} $,and $V_{2,k_{y0}}=V_{2} $ with
$\vec{k}_0$ being the position of the Weyl point.

The effective two dimensional Hamiltonian is then 
obtained by projecting the Hamiltonian onto these eigenvalues,  
$[\mathcal{H}_{\rm eff}]_{i,j} = \langle v_i|  \mathcal{H}_{1} (\delta
{\bf k}) | v_j \rangle $ with $i,j=1,2$, giving rise to 
\begin{widetext}
\begin{align}
\mathcal{H}_{\rm eff} (\delta {\bf k}) =-h_{0,\vec{k}_0}\sigma_z
 - V_{2,k_{y0}}\delta k_y \sigma_x +
 \frac{V_{1,k_{x0}}+\sqrt{V^2_{1,k_{x0}}+V^2_{3,k_{z0}}}}{V^2_{1,k_{x0}}+V^2_{3,k_{z0}}+V_{1,k_{x0}}\sqrt{V^2_{1,k_{x0}}+V^2_{3,k_{z0}}}} 
\left[ V_{1,k_{x0}} \tilde{V}_{1,k_{x0}} \delta k_x + V_{3,k_{z0}} \tilde{V}_{3,k_{z0}} \delta k_z\right] \sigma_y,
\label{Eq:kp}
\end{align}
where $h_{0,\vec{k}_0}=\frac{1}{2} (1+\alpha) [2 t_x \sin {k_{x0} \delta k_x}+2 t_z \sin {k_{z0} \delta k_z}]$,
and  $\tilde{V}_{1(3),k_{x(z)0}}=V_{1(3)}\cos k_{x(z)0} $.

For simplicity, we express the Hamiltonian as
 \begin{align}
 \mathcal{H}_{\rm eff} (\delta {\bf k})=(A \delta k_x + B \delta k_z )\sigma_z
 + (C \delta k_x + D \delta k_z )\sigma_y
 +E \delta  k_y  \sigma_x ,
 \end{align}
 where 
\begin{align} 
&{A}=-\frac{1+\alpha}{2}(2 t_x \sin k_{x0}), \quad {B}=-\frac{1+\alpha}{2}(2 t_z \sin k_{z0})%\notag\\
\quad{C}= \frac{V_{1,k_{x0}}+\sqrt{V^2_{1,k_{x0}}+V^2_{3,k_{z0}}}}{V^2_{1,k_{x0}}+V^2_{3,k_{z0}}+V_{1,k_{x0}}\sqrt{V^2_{1,k_{x0}}+V^2_{3,k_{z0}}}} V_{1,k_{x0}} \tilde{V}_{1,k_{x0}} , \notag\\
& {D}= \frac{V_{1,k_{x0}}+\sqrt{V^2_{1,k_{x0}}+V^2_{3,k_{z0}}}}{V^2_{1,k_{x0}}+V^2_{3,k_{z0}}+V_{1,k_{x0}}\sqrt{V^2_{1,k_{x0}}+V^2_{3,k_{z0}}}} V_{3,k_{z0}} \tilde{V}_{3,k_{z0}},%\notag\\
\quad {E}=-V_{2,k_{y0}}.
\end{align}
\end{widetext}

\section{Topological invariance of the Weyl points---Berry curvature}
\label{App:C}
The topological invariance of the Weyl  points 
is the Berry curvature computed from a two-dimensional surface 
encircling the Weyl point. The definition of the Berry curvature
is
\begin{align}
C= \frac{i}{2 \pi}\sum_{\alpha \in {\rm occ.}}\int d^2 k \langle \partial_{k_i}u_\alpha| \partial_{k_j}u_\alpha\rangle-(k_i \leftrightarrow k_j),
\end{align}
where $u_\alpha$ are the occupied bands and the two dimensional integral is a closed surface around one Weyl point.

%The effective two-dimensional Hamiltonian around the Weyl point is
Now we compute the Berry curvature around the Weyl point by using Eq. (\ref{Eq:kp}).
Without loss of generality, we define $\tilde{k}_x= A \delta k_x+ B \delta k_z $,  
$\tilde{k}_z= C \delta k_x+D \delta k_z$, and $\tilde{k}_y= E \delta k_y$.

We now choose a fixed radius $R$ around the Weyl point with $R^2=\sum_{i=x,y,z} \tilde{k}_i^2$. 
The occupied band with energy $-R$ is
\begin{align}
u_-(\theta,\phi) = \left(\begin{array}{c}-\sin \frac{\theta}{2} e^{- i \phi} \\ \cos \frac{\theta}{2} \end{array}\right),
\end{align}
where we parameterize $\tilde{k}_{z}=R \cos \theta$, $\tilde{k}_{x}=R \sin \theta \cos \phi$, and 
$\tilde{k}_{y}=R \sin \theta \sin \phi$. 
%[Warning: $\tilde{k}_{z}$ is not orthogonal to $\tilde{k}_{x}$, the parameterization here is not an ordinary spherical 
%coordinate centered at the Weyl point].

The only non-vanishing component of the Berry connection  is
\begin{align}
A_{\phi}=\frac{1}{R\sin \theta} \langle u_- (\theta,\phi) |\partial_{\phi} u_- (\theta,\phi) \rangle=\frac{i}{2 R} \cot \theta.
\end{align}
The Berry curvature around the Weyl point is then
\begin{align}
C=\frac{i}{2 \pi} \int_{\rm sphere} d s ( \nabla \times \vec{A})
=\frac{i}{2 \pi} \int R^2 \sin \theta d \theta d \phi \frac{-i}{2 R^2} =1.
\end{align}
The Berry curvature of teh 
the other time-reversal related Weyl points at $-\vec{k}_0$,
is $C=-1$.

\section{Fermi arc states in the effective Hamiltonian }
\label{App:D}
We analyze the Fermi arc state from the effective two-dimensional Hamiltonian 
in Eq. (\ref{Eq:kp}). In the presence of $(010)$ surface, the effective Hamiltonian around the Weyl point is expressed as
\begin{align}
\mathcal{H}_{\rm eff.}=\left(\begin{array}{cc}A \delta k_x+{B} \delta k_z & -i{C} \delta k_x -i {D} \delta k_z -  i{E} \partial_y \\
 iC\delta k_x + i{D} \delta k_z-i{E} \partial_y & -{A}\delta k_x-{B} \delta k_z\end{array}\right).
\end{align}
We consider a cylindrical 
surface surrounding the Weyl point with radius $k_0$. %as shown in Fig. \ref{Fig_8}(a). 
The effective Hamiltonian becomes
\begin{widetext}
\begin{align}
\mathcal{H}_{\rm eff.}=
\left(\begin{array}{cc}A k_0 \cos \theta +{B} k_0 \sin \theta & -ik_0 \sqrt{{C}^2+{D}^2}\cos (\theta-\phi)- i{E} \partial_y \\
ik_0 \sqrt{{C}^2+{D}^2}\cos (\theta-\phi) - i E \partial_y & -{A} k_0 \cos \theta-{B} k_0 \sin \theta
\end{array}\right),
\end{align}
where $\phi=\cos^{-1} \frac{C}{\sqrt{C^2+D^2}}$, $\delta k_x= k_0 \cos \theta$, and $\delta k_z = k_0 \sin \theta$.

There are two boundary states on this cylinder surrounding the Weyl point
\begin{align}
&u_{y>0}=\left(\begin{array}{c}1 \\0\end{array}\right) \sqrt{2 \kappa} e^{- \kappa y} \quad {\rm with} \quad E_R(\theta)={A}k_0 \cos \theta + B k_0 \sin \theta,  \\ \notag
&u_{y<0}=\left(\begin{array}{c}0 \\1\end{array}\right) \sqrt{2 \kappa} e^{ \kappa y} \quad {\rm with} \quad E_L(\theta)=-{A} k_0 \cos \theta - B k_0 \sin \theta,
\end{align}
where $\kappa=-\frac{k_0}{E}\sqrt{{C}^2 + {D}^2} \cos (\theta-\phi)>0$. 
These boundary states are the origin of the Fermi arc states. 
%The zero energy of the bound states is at $\theta = \cot ^{-1} [-\frac{A}{B}]$ for all radius $k_0$. 
%The spectrum as a function of $\theta$ of two boundary states are shown in Fig. \ref{Fig_8}(b).

%\begin{figure}[htbp]
%\centering
%\includegraphics[height=4 cm] {Fig8.pdf}
%\caption{Bulk nodal rings along $k_z$ direction. The surface flat band emerges on $(011)$ surface bounded by the rings projected on the surface is shown schematically. }
%\label{Fig_8}
%\end{figure}

%Now we examine the effect of Kondo breakdown on the Fermi arc states.
%The surface Kondo breakdown is the removal of the hybridization terms on the surface,
%which is equivalent to introduce a $\delta \mathcal{H}({\bf k})=-\delta(y)( -i E \partial_y \sigma_x)$.
%By using the second order perturbation theory, the energy of the surface band will be shifted by
%\begin{align}
%\delta E &= -\sqrt{2\kappa}((\frac{-AB }{(A^2+B^2)}+\frac{-AC}{(A^2+C^2)}) \alpha_3+\frac{A(C-B)\alpha_1}{\sqrt{(A^2+B^2)(A^2+C^2)}}) \quad {\rm for} \quad u_{y>0}  \notag \\
%&=-\sqrt{2\kappa}((\frac{-AB }{(A^2+B^2)}+\frac{-AC}{(A^2+C^2)}) \alpha_3-\frac{A(C-B)\alpha_1}{\sqrt{(A^2+B^2)(A^2+C^2)}}) \quad {\rm for} \quad u_{y<0}.
%\end{align}
%The surface spectrum in the presence of Kondo breakdown in shown in Fig. \ref{Fig_8}(b) by dashed lines.
\end{widetext}

%\subsection{Topological invariant of the nodal rings---winding number}
\section{Time-reversal symmetric nodal ring semimetallic phase}
\label{App:E}

 \begin{figure*}[htbp]
\centering
\includegraphics[height=6 cm] {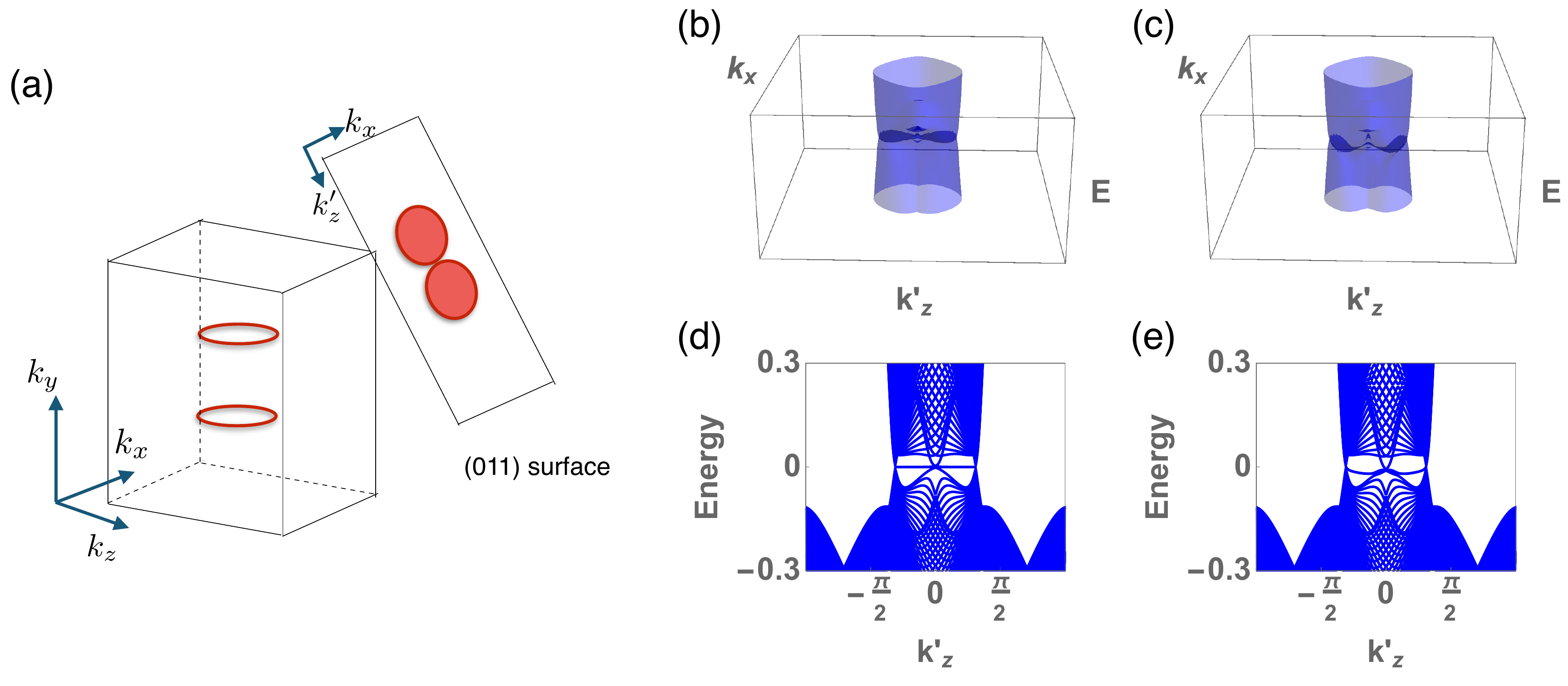}
\caption{(a) Schematic plot of the bulk nodal rings 
centered at $k_y$ axis. On the $(011)$ surface, the surface flat bands will emerge inside the circles which
are the projection of the bulk nodal rings. The surface spectrum as a function of $(k_x, k'_z=\frac{1}{\sqrt{2}}(k_z-k_y))$
from the Hamiltonian Eq. (\ref{Ham}) with $(t_x,t_y,t_z,\mu,\alpha,W_x, W_y,W_z,W_0)=(2,1,1,-6.5,-0.1,2,2,2,1.8)$. 
(b) in the absence of surface Kondo breakdown and (c) in the presence of surface Kondo breakdown.
The surface spectrum as a function of $k'_z=\frac{1}{\sqrt{2}}(k_z-k_y)$ at $k_x=0$,
(d) in the absence of surface Kondo breakdown and (e) in the presence of surface Kondo breakdown.}
\label{Fig_3}
\end{figure*}

The Hamiltonian of hWSMs with non-vanishing $W_0$ 
can be expressed as 
\begin{align}
\mathcal{H}({\bf k})= \mathcal{H}_0({\bf k}) + \mathcal{H}_1({\bf k}),
\end{align} 
where $ \mathcal{H}_0({\bf k}) =\frac{1}{2}(\epsilon_c({\bf k})+\epsilon_f({\bf k}) ) \sigma_{0}\tau_{0}$ and
$\mathcal{H}_1({\bf k})=\frac{1}{2}(\epsilon_c({\bf k})-\epsilon_f({\bf k}) ) \sigma_{0}\tau_{3}+ \sum_{i=1,2,3} V_i  \sin k_i \sigma_i\tau_1
 +W_0 \tau_1.$%+W_2\tau_2\sigma_2$.
%In the absence of inversion breaking $W_2$ term, 
$\mathcal{H}_1({\bf k})$ has a chiral symmetry, $\mathcal{S}^{-1}\mathcal{H}({\bf k})\mathcal{S}= - \mathcal{H}({\bf k})$,
where $\mathcal{S} = \tau_2$.
In the presence of chiral symmetry, one can off-block diagonalize $\mathcal{H}_1({\bf k})$ by a unitary transformation,
$\mathcal{V}^{\dagger } \mathcal{H}_1({\bf k})\mathcal{V}=\tilde{\mathcal{H}_1}({\bf k})$, where
\begin{align}
\mathcal{V}= \frac{1}{\sqrt{2}}\left(\begin{array}{cc}\mathbb{I}_{2 \times 2} & i \mathbb{I}_{2 \times 2} \\i \mathbb{I}_{2 \times 2} & \mathbb{I}_{2 \times 2}\end{array}\right),
\quad 
\tilde{\mathcal{H}}({\bf k})_1=\left(\begin{array}{cc}0 & D({\bf k}) \\D^\dagger({\bf k}) & 0\end{array}\right),
\end{align}
with $D({\bf k})= i (\epsilon_c({\bf k })-\epsilon_f ({\bf k})) +2 \sum_i V_i\sin k_{i} \sigma_i+m_0$.
The eigenvectors of $\tilde{\mathcal{H}}({\bf k})$ satisfy
\begin{align}
\left(\begin{array}{cc}0 & D({\bf k}) \\D^\dagger({\bf k}) & 0\end{array}\right) \left(\begin{array}{c}\chi^{\pm} ({\bf k}) \\\eta^{\pm} ({\bf k})\end{array}\right)
=\pm\lambda ({\bf k})  \left(\begin{array}{c}\chi^{\pm} ({\bf k}) \\\eta^{\pm} ({\bf k})\end{array}\right).
\label{Eq.7}
\end{align}
These eigenvectors are also the eigenvector of $\mathcal{H}_0({\bf
k})$ and they determine the topological invariant. 
We pick $\chi^{\pm}({\bf k})=\frac{1}{\sqrt{2}}u({\bf k})$, Then Eq. (\ref{Eq.7}) leads to 
$\eta^{\pm}({\bf k})=\pm\frac{1}{\sqrt{2}}\frac{1}{\lambda({\bf k})}D^\dagger({\bf k}) u({\bf k})$.
The flat band Hamiltonian $\mathcal{Q}({\bf k})$can be obtained from the projector
\begin{align}
\mathcal{Q}({\bf k})=&\mathbb{I}-2 \sum_{\alpha \in {\rm occ.}}|u_{\alpha} ({\bf k}) \rangle\langle  u_{\alpha}({\bf k})|   \notag\\
=&\mathbb{I}-\left(\begin{array}{c}u({\bf k}) \\-\frac{1}{\lambda({\bf k})}D^\dagger({\bf k}) u({\bf k})\end{array}\right)\left(\begin{array}{cc}u^\dagger({\bf k}) & -\frac{1}{\lambda({\bf k})}u^\dagger({\bf k})D({\bf k})\end{array}\right) \notag \\
=&\frac{1}{\lambda({\bf k})}\left(\begin{array}{cc}0 & u({\bf k }) u^\dagger({\bf k }) D({\bf k}) \\D^\dagger({\bf k}) u ({\bf k}) u^\dagger ({\bf k}) & 0\end{array}\right)  \notag\\
=&\left(\begin{array}{cc}0 & q({\bf k}) \\q^\dagger({\bf k}) & 0\end{array}\right).
\end{align}

The topological invariance of the nodal ring is characterized by a winding number of a one-dimensional loop
encircling the ring. The winding number can be calculated from the
q-matrix integral
\begin{align}
\nu=\frac{1}{2 \pi i} \oint_\mathcal{L} d{\bf k} {\rm Tr} [q^{-1} ({\bf k})\partial_{\bf k} q ({\bf k})].
\end{align}
In our model, the winding number of the nodal rings is $\nu=\pm 1$,
which leads to surface flat bands bounded by the nodal rings projected on the surface Brillouin zone\cite{Burkov2011,Matsuura2013}. 
As shown schematically in Fig. \ref{Fig_3}(a), two nodal rings are centered along $k_y$ axis.
On $(011)$ surface, the flat band surface states emerge inside the bulk rings projected on the $(011)$ surface Brillouin zone.

Finally, we have investigated the Kondo breakdown on these surface
flat bands, summarizing the results in \ref{Fig_3}. 
In the absence of the surface Kondo breakdown,
the surface flat bands emerge on $(011)$ surface
(Fig. \ref{Fig_3}(b) and (d)). . 
In the presence of the surface Kondo breakdown, the surface flat bands sink beneath the Fermi sea
as shown in Fig. \ref{Fig_3}(c) and (e).

%\bibliographystyle{apsrev4-1}
%\bibliography{references}

\bibliographystyle{apsrev4-1}
\bibliography{references}

\end{document}